\documentclass[sigconf,nonacm]{acmart}
\usepackage{graphicx} % Required for inserting images
\usepackage{paralist}
\usepackage{cleveref}
\usepackage{tikz}
\usetikzlibrary{positioning}

\title{Property-Driven Synthetic Data Engineering for Data-Scarce Software Systems: Reflections from the Breast Cancer Domain} 
\author{Aurora Francesca Zanenga}
\email{aurora.zanenga@unibg.it}
\orcid{0009-0008-0655-8335}
\affiliation{%
    \institution{University of Bergamo}
    \city{Bergamo}
    \country{Italy}
}

\author{Andrea Bombarda}
\email{andrea.bombarda@unibg.it}
\orcid{0000-0003-4244-9319}
\affiliation{%
    \institution{University of Bergamo}
    \city{Bergamo}
    \country{Italy}
}

\author{Marsha Chechik}
\email{chechik@cs.toronto.edu}
\orcid{0000-0002-6301-3517}
\affiliation{%
    \institution{University of Toronto}
    \city{Toronto}
    \country{Canada}
}

\author{Saverio D'Amico}
\email{saverio.damico@humanitas.it}
\orcid{0000-0003-1877-8497}
\affiliation{%
    \institution{Humanitas Clinical and Research Center, IRCCS}
    \city{Rozzano}
    \country{Italy}
}

\author{Rita De Sanctis}
\email{rdesanctis@asst-pg23.it}
\orcid{0000-0003-3202-1933}
\affiliation{%
    \institution{ASST Papa Giovanni XXIII}
    \city{Bergamo}
    \country{Italy}
}

\author{Alberto Zambelli}
\email{azambelli@asst-pg23.it}
\orcid{0000-0002-1374-1831}
\affiliation{%
    \institution{ASST Papa Giovanni XXIII}
    \city{Bergamo}
    \country{Italy}
}

\author{Claudio Menghi}
\email{claudio.menghi@unibg.it}
\orcid{0000-0001-5303-8481}
\affiliation{%
    \institution{University of Bergamo}
    \city{Bergamo}
    \country{Italy}
}
\affiliation{%
    \institution{McMaster University}
    \city{Hamilton}
    \state{ON}
    \country{Canada}
}

\begin{document}

\begin{abstract}

%Modern software applications are data-driven. 
%Data are used to support software reasoning and decision-making.
%Despite their importance, many software applications lack sufficient data.
%Engineering software for these applications poses fundamental challenges for the software engineering community.
%This paper reflects on our experience engineering medical software in collaboration with oncologists from a major hospital specializing in breast cancer treatment.
%Our findings are derived from the analysis of data from the intraoperative radiotherapy (IORT) treatment.
%Unlike many other domains where data are widely available, we present unexpected insights derived from a limited amount of data on specific population groups, who often benefit most from early diagnosis and treatment (e.g., young female patients).
%This paper challenges existing research outcomes, encourages the community to reconsider current directions, and motivates new avenues for future work.

Modern software systems increasingly depend on data for analysis, prediction, testing, and decision-making. Yet many important domains, including medicine, safety-critical systems, and regulated industries, lack abundant, shareable, or representative data. Synthetic data generation is often proposed as a remedy, but our experience engineering software for intraoperative radiotherapy (IORT) in breast cancer treatment suggests that synthetic data shifts rather than solves the central engineering problem. 
The key challenge becomes deciding which properties synthetic data must preserve, how these properties should be elicited from stakeholders, how they can be validated under privacy constraints, and how they evolve. We call this problem \emph{property-driven synthetic data engineering}. 
Drawing on a collaboration with oncologists and preliminary experiments with a sensitive IORT dataset, we identify challenges in requirements, validation, privacy, and pipeline evolution. We argue that automated software engineering research should develop methods and tools for eliciting, formalizing, checking, and evolving validity properties for synthetic data in data-scarce software systems.
    
\end{abstract}




\maketitle

\section{Introduction}
Data are driving modern software applications.
%They are necessary to perform tasks such as prediction and classification, and the growing use of artificial intelligence is increasing the need for data.
Tasks such as prediction, classification, testing, and decision-making rely on the availability of large, accessible datasets, and the increasing adoption of artificial intelligence further reinforces this assumption.
%Despite significant amounts of data being available in many applications (e.g., social media platforms, streaming websites), there are application domains where data are difficult to retrieve, extremely valuable, and contain sensitive information.
However, in many safety-critical and privacy-sensitive domains, this assumption does not hold. 
In domains such as medicine and regulated industries, data are often scarce, difficult to access, and subject to strict confidentiality constraints. 
As a result, software engineering (SE) practices that implicitly rely on abundant data become difficult to apply.
This is the case with our partner, a large hospital with more than one million outpatient services and 100,000 emergency department visits.
Despite the large number of patients, in many medical scenarios, the number of patients affected by certain pathologies is (fortunately) limited, and the data contain sensitive health information.
For example, in the case of breast cancer, while the overall number of cases is significant (approximately 2.3 million new cases each year~\cite{world2024global}), the therapeutic landscape~\cite{cardoso2024breast} is rapidly evolving, and the number of patients within specific subpopulations (e.g., young women or minority groups) may be very small or even absent.
%
%For example, consider a specific oncological pathology like breast cancer.
%Breast cancer has approximately 2.3 million new cases each year~\cite{world2024global}.
%It is the second cause of cancer-related mortality~\cite{bray2024global} and a rapidly evolving therapeutic landscape~\cite{cardoso2024breast}.
%Our partner hospital has approximately 1000 accesses related to this pathology each year.
%However, the number of accesses related to a particular segment of the population (e.g., young women) is (luckily) very low, and for certain population groups (e.g., specific minorities), completely absent.
Consequently, engineering software relying on such data introduces specific challenges that contrast with mainstream SE practices based on abundant and shareable datasets.

To address data scarcity, Synthetic Data Generation (SDG) techniques create artificial datasets that mimic properties (e.g., patterns, distributions, and correlations) of existing datasets. 
%It is particularly relevant in data-driven domains where access to high-quality data is limited or constrained.
These techniques are widely proposed as a remedy in contexts where data are limited or cannot be shared due to privacy constraints~\cite{10.1007/978-3-032-22774-4_5}.
%These techniques are also beneficial in contexts where data privacy and sensitive information are present, like the medical domain, where data availability is often restricted due to ethical, legal, and privacy concerns~\cite{10.1007/978-3-032-22774-4_5}.
For example, clinical datasets typically contain personal and identifiable information, making their sharing and reuse across institutions challenging. 
However, our experience suggests that synthetic data do not eliminate the core problem; rather, they shift it. 
The central challenge becomes determining which properties synthetic data must preserve, how these properties can be elicited from stakeholders, how they can be validated in the absence of ground truth, and how they evolve over time under changing requirements and constraints.
%
%As a consequence, the development of synthetic datasets that accurately preserve the statistical properties of real data while preventing the disclosure of sensitive information has become increasingly important.
%Consequently, SDG gained significant attention in software engineering, where it is used to support requirement analysis, testing, benchmarking, and empirical studies when real-world data is scarce, proprietary, or difficult to obtain~\cite{Soltana2017,Skopik2014}.
%However, synthetic data do not solve the validation problem by themselves; they shift it from data availability to reasoning about which properties must be preserved under privacy constraints and domain-specific requirements.
We argue that data scarcity fundamentally breaks standard SE assumptions, requiring a shift from data-driven validation to property-driven synthesis and validation.
In such settings, developers cannot rely on large datasets for validation, testing, or model training. Instead, they must reason in terms of properties, constraints, and synthetic approximations. This shift introduces new challenges in requirements engineering, validation, and system design that are not addressed by existing methodologies.
To address this gap, this paper proposes \emph{property-driven synthetic data engineering} as a research agenda for automated SE, in which the central artifacts are stakeholder-specific validity properties that must be elicited, formalized, checked, and evolved.

%This work is motivated by a collaboration with oncologists from a major hospital specializing in breast cancer treatment. 
%Breast cancer represents one of the most prevalent and impactful diseases worldwide~\cite{ARNOLD202215}: according to global statistics, approximately one in eight women (around 12–13\%) will develop breast cancer during their lifetime. 
%Timely diagnosis and treatment significantly impact the outcomes of treatments.
%Specifically, early breast cancer (EBC) can be cured with multi-modal treatment (although not without side effects).
%We focus on a specific treatment, namely intraoperative radiotherapy (IORT), which delivers (immediately after the tumor is removed) a high-concentration dose of radiation~\cite{zazzetti2025longitudinal}.
We expose this problem through a case study in breast cancer treatment, namely intraoperative radiotherapy (IORT), which delivers (immediately after the tumor is removed) a high-concentration dose of radiation~\cite{zazzetti2025longitudinal}.
While highly valuable for clinical research, the data from our partner hospital cannot be freely shared due to strict privacy constraints; at the same time, enabling access to those data would significantly benefit the research community, allowing for more extensive experimentation and validation of medical and computational approaches.
To overcome these challenges, our project envisions the application of SDG solutions and highlights the need to treat SDG not as a pre-processing step, but as an engineering process centered on validity properties.

%In particular, this paper proposes \emph{property-driven synthetic data engineering} research agenda for automated software engineering.
In this paper, we reflect on our collaboration, preliminary experimentation, challenges we faced, and lessons we learned.
%Rather than focusing only on data generation, we highlight how validation becomes challenging in data-scarce and privacy-sensitive settings, where developers must reason about which properties of the data should be preserved.
We argue that SDG for software systems should be treated not as a data pre-processing technique, but as an engineering process whose central artifacts are stakeholder-specific validity properties.
In data-scarce and privacy-constrained settings, developers must elicit, formalize, check, trade off, and evolve properties such as statistical fidelity, clinical plausibility, privacy protection, and task-specific utility. 
%In data-scarce and privacy-constrained settings, developers cannot validate systems by relying on abundant ground-truth data. 
%Instead, they must elicit, formalize, check, trade off, and evolve properties such as statistical fidelity, clinical plausibility, privacy protection, and task-specific utility. We call this problem \emph{property-driven synthetic data engineering} and identify it as an emerging challenge for automated software engineering.
We call this problem \emph{property-driven synthetic data engineering} and identify it as an emerging challenge for automated SE.
%In particular, we focus on software engineering challenges related to requirements, validation, stakeholder needs, and the evolution of synthetic data pipelines.
%We show that statistical similarity alone does not necessarily imply clinical validity, and that different requirements (e.g., statistical fidelity, clinical plausibility, and privacy) may lead to trade-offs. \aurora{Suggerimento di Marsha, da inserire solo se riusciamo a fare in tempo la parte di validazione clinica e se effettivamente risulta essere vera questa cosa}
In particular, the following are the contributions of this paper:
\begin{compactitem}
    \item It identifies data absence as a SE problem that affects requirements, validation, testing, and system evolution in data-driven software.
    \item It introduces a \emph{property-driven} view of synthetic data engineering, in which synthetic datasets are evaluated against stakeholder-specific properties rather than only against global similarity metrics.
    \item It reports early lessons from an IORT breast-cancer case study and derives a research agenda for automated support for property elicitation, property checking, trade-off analysis, and pipeline evolution.
\end{compactitem}
%We present insights and challenges observed in the early phases of this collaboration.
%Based on our experience, we discuss key software engineering challenges for domains characterized by limited and sensitive data, and outline directions for future work.

This paper is structured as follows. 
\Cref{sec:problem} presents our research problem.
\Cref{sec:preliminary} presents our preliminary experimentation and results.
\Cref{sec:challenges} presents challenges and lessons learned.
\Cref{sec:related} presents related work.
\Cref{sec:conclusions} presents our conclusions.

\section{Research Problem and Context}
\label{sec:problem}
This work is based on a collaboration with the oncologists of a major hospital specializing in breast cancer treatment.
We aim to support physicians by designing a software system that that supports clinical data-analysis and decision-support pipeline, allowing oncologists to explore IORT outcomes, evaluate recurrence-related factors, and test analysis workflows without exposing identifiable patient data. 
Our effort in analyzing SDG is intended not to replace clinical evidence, but to support software development, testing, benchmarking, and controlled sharing of analysis pipelines.
%We aim to support physicians by designing software that supports data analysis and management for the intraoperative radiotherapy (IORT) treatment. 
We received a dataset containing clinical and treatment-related variables collected during IORT procedures. 
The dataset consists of 1000  patients described by 64 variables (before data cleaning). 
This is a considerably large dataset for this domain, both in terms of samples (not all the patients affected by breast cancer are treated with IORT) and features (64). 
This dataset contains sensitive information (e.g., personal information that are highly confidential).
The size of the dataset and the sensitivity of the contained information pose significant challenges for engineering software applications that can effectively support physicians.
On the one hand, despite being a considerably large dataset for this domain, the limited number of samples hampers the effective use of existing AI solutions that need a significant amount of data to ensure the reliability of the results produced by the analysis.
On the other hand, considering the sensitive nature of the data, the distribution of the dataset is forbidden. 
These constraints make it difficult not only to build models but also to define how results should be validated.

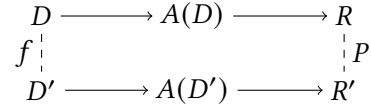
\begin{figure}[t]
\centering
\begin{tikzpicture}[
    node distance=2cm and 0.3cm,
    every node/.style={font=\Large}
]

% Nodes
\node (D) {$D$};
\node (AD) [right of=D] {$A(D)$};
\node (R) [right of=AD] {$R$};

\node (Dprime) [below=.5cm of D] {$D'$};
\node (ADprime) [right of=Dprime] {$A(D')$};
\node (Rprime) [right of=ADprime] {$R'$};

% Arrows
\draw[->] (D) -- (AD);
\draw[->] (AD) -- (R);

\draw[->] (Dprime) -- (ADprime);
\draw[->] (ADprime) -- (Rprime);

\draw[dashed] (D) -- node[left] {$f$} (Dprime);

\draw[dashed] (R) -- node[right] {$P$} (Rprime);

% Brace + condition
%\draw[
%    decorate,
%    decoration={brace,amplitude=6pt}
%] ($(R)+(0.4,0.2)$) -- ($(Rprime)+(0.4,-0.2)$)
%node[midway,right=10pt] {$|R - R'| \leq \epsilon$};

\end{tikzpicture}
\caption{Synthetic data generation: A high-level overview.}
\label{fig:formalization}
\vspace{-.5em}
\end{figure}

SDG is a technique that may address these challenges. 
\Cref{fig:formalization} outlines our formalization of the \emph{property-driven synthetic data} generation for software applications.
Consider a scenario in which a (limited) set of data ($D$) leads to the results $R$ when processed by an analysis ($A$).
Examples of such an analysis are survival analysis, risk prediction, patient stratification, and testing of a clinical decision-support system.
SDG concerns the development of a procedure $f$ that generates a set of data ($D^\prime$) which leads to the results $R^\prime$ when processed by the same analysis ($A$).
Examples of results are Kaplan-Meier survival curves, estimation of Cox model coefficients, model predictions, or test coverage of patient subgroups.
We argue that SDG should ensure that $R^\prime$ satisfies a set of properties $P$, e.g., distributional similarity, clinically legal value ranges, plausible treatment sequences, low privacy leakage, or utility for downstream tasks. 
Such properties are stakeholder-dependent and potentially conflicting.
%Our experience suggests that there may be, for example, a trade-off between privacy preservation and task-specific utility.
For example, clinicians may require clinical plausibility, data scientists statistical fidelity, and privacy officers strong disclosure protection. 
As a result, satisfying one property (e.g., statistical similarity) may hamper the satisfaction of others (e.g., privacy preservation), requiring specific trade-offs.
%For instance, synthetic data generation enables addressing limitations in the size of the data by generating (large) synthetic datasets preserving the clinical and statistical properties of the original dataset (e.g., requiring the difference between the results $R$ and $R'$ to be lower than an acceptable threshold $\epsilon$). 
%However, a synthetic dataset may preserve correlations observed in the original data, while producing clinically implausible (e.g., out-of-scale) values or unrealistic combinations of variables.
%Additionally, it may generate new data that exposes sensitive information from the patients.
For instance, synthetic data may preserve correlations observed in the original data while producing clinically implausible values or unrealistic combinations of variables, or may expose sensitive information.
This highlights that the SE challenge is not only generating synthetic data, but understanding which properties matter for a given task and how they can be validated in the absence of ground truth.
 
% For this reason, the goal of this work is to investigate whether synthetic data generation techniques can be effectively applied in this context to produce datasets that preserve the key statistical and clinical characteristics of the original data while ensuring privacy preservation.

%The primary objective of this study is to generate a synthetic version of the IORT dataset that can be safely shared and used by external researchers \aurora{controllare questa cosa}.

%\begin{figure*}[t]
%    \centering
%    \begin{subfigure}[b]{0.425\textwidth}
%        \centering
 %       \includegraphics[width=\linewidth]{figures/corr_matrix_IORT.png}
 %       \caption{Correlation matrix for the original dataset IORT}
 %       \label{fig:matrixIORT}
 %   \end{subfigure}
 %   \hfill
 %   \begin{subfigure}[b]{0.485\textwidth}
 %       \centering
 %       \includegraphics[width=\linewidth]{figures/corr_matrix_IORT_Synth.png}
 %       \caption{Correlation matrix for the generated dataset IORT}
 %       \label{fig:matrixIORTs}
 %   \end{subfigure}   
 %   \caption{Comparison between the correlation matrix of the original dataset and the generated dataset.}
 %   \label{fig:comparison}
%\end{figure*}

\section{Preliminary Experimentation}
\label{sec:preliminary}
In our preliminary experimentation, we conducted three activities:  (a)~cleaning our dataset, (b)~selecting and applying an appropriate SDG technique, and (c)~analyzing our results.

\emph{Dataset Cleaning}. We manually cleaned our dataset to improve data quality. 
We removed columns that contained empty values for more than 80\% of the rows (e.g., ``Previous Tumor Date'', ``Chemotherapy Start Date'', ``Days from IORT to Chemotherapy Start'', ``Percentage Difference Between Administered and Prescribed Dose'', ``Mastectomy Date''). 
Since those columns had a high percentage of missing values, they were considered not relevant in most of the cases. 
Moreover, due to the large amount of missing data, it was not possible to reliably model their correlations with the other variables. 
Also, such variables could negatively affect the quality of the SDG process.
We fixed typos in entry values (e.g., the word ``fulfilled'' was often written in different ways, such as ``full filled'' or ``fulfilled''), standardized these entries, and removed rows containing values that did not fall within the set of valid values for a given column (e.g., when the allowed values were 0, 1, or 2, but a value of 5 was present). 
After pre-processing, the final dataset contained 709 patients and 58 variables (down from 64).
We discussed our changes with oncologists to ensure the soundness of our cleaning activity (e.g., we verified that the columns that had been removed did not contain any relevant information).
Such a process highlighted a critical challenge: data are not always ready to be used.

\emph{Synthetic Data Generation}.
We reviewed existing SDG techniques from a recent work~\cite{shi2025synthetic} that could be suitable for our context and have been applied to the medical domain.
We selected TabDDPM~\cite{pmlr-v202-kotelnikov23a}, CTGAN~\cite{synthetic_healthcare_2025}, Gaussian Copula~\cite{9346329}, CopulaGAN~\cite{app14145975}, and Tabular Variational Autoencoder (TVAE)~\cite{10101315} since they can process tabular data and have been applied in similar contexts.

\emph{Analyzing our results}. After some preliminary experiments, we evaluated the quality of the generated data by analyzing both the relationships between variables and their individual distributions. 
First, we analyzed the correlation matrices.
The correlation matrices enable us to assess whether the relationships among variables were preserved.
In addition, we analyzed the distributions of synthesized variables, which closely match those of the original variables. 
While these results suggest that these methods can preserve the statistical properties of the data, they do not necessarily guarantee that the generated data are clinically plausible.
This observation further motivates our formalization of the SDG task for SE, where different properties may be considered by different stakeholders and different SDG approaches may be used.
Additionally, our analysis showed that different evaluation views lead to different conclusions, so generator selection is a requirements- and validation-related problem, not only an ML model-selection problem.

We observed that TVAE~\cite{10101315}, a deep learning generative model built on a Variational Autoencoder (VAE), achieved better results than the other approaches, given the characteristics of our dataset (i.e., a limited number of samples and a relatively high number of variables). 
Specifically, for TVAE, the two correlation matrices show a high similarity, showing that the model can capture and reproduce the main relationships between variables.
%TVAE ~\cite{10101315} is a deep learning generative model built on the Variational Autoencoder (VAE) architecture. 
%An autoencoder is a neural network composed of two main components: an encoder and a decoder. The encoder takes the input data (in our case, 58 variables) and maps it into a lower-dimensional latent space. 
%The decoder reconstructs the data from this latent representation, producing an output with the same structure as the original input. 
%By learning the compressed representation, the model can capture the underlying patterns of the data and generate new synthetic samples that resemble the original dataset.

% For this reason, we focused our discussion on this model.

%\begin{figure*}[t]
%    \centering
%    \includegraphics[width=0.8\textwidth]{figures/plot.png}
%    \caption{Plot Comparison.}
%    \label{fig:plot}
%\end{figure*}

%\Cref{fig:matrixIORT} shows the correlation matrix of the original dataset, while \Cref{fig:matrixIORTs} presents the correlation matrix of the synthetic data generated using TVAE. 

%As we can see in Figure \ref{fig:plot} compares selected variable distributions from the original dataset (shown in blue) with those from the synthetic dataset (shown in orange). The distributions appear to be closely aligned, indicating that the model is able to approximate the statistical properties of the original data.

\begin{figure}[t]
        \includegraphics[width=\linewidth]{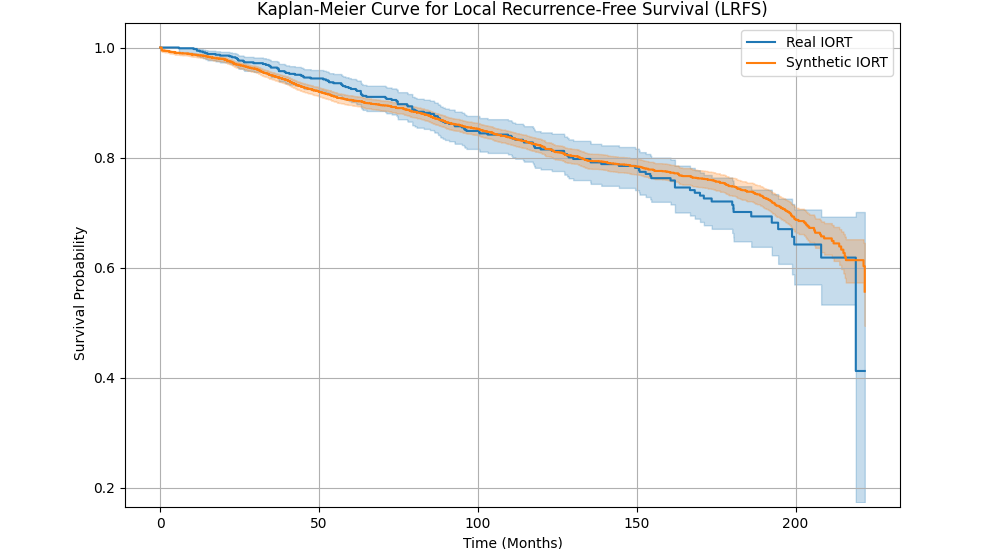}
        \vspace{-2em}
        \caption{Survival Curves: IORT vs Synthetic IORT}
                \label{fig:LRFS}
        \vspace{-.7em}
\end{figure}

%\subsection{Clinical Validation}
%\label{sec:clinical}
To analyze the clinical validity of the synthetic IORT dataset, we studied the Local Recurrence-Free Survival (LRFS).
We compared the real and synthetic data using four steps:
(I) Global Kaplan-Meier curves, (II) Stratified Kaplan-Meier curves, (III) Univariate Cox models, and (IV) Multivariate Cox models.
These analyses enable the verification of patient survival patterns and clinical relationships.
For example, we report the results of the Kaplan-Meier analysis, a method used to estimate survival over time. 
We compared the global Kaplan-Meier curves~\cite{doi:10.1016/j.otohns.2010.05.007} between the real and synthetic data.
The curves in \Cref{fig:LRFS} showed a very similar trend over most of the follow-up period. 
In both datasets, the probability of remaining free from local recurrence gradually decreased over time.
Some differences appeared at the end of the curves, where the number of patients was low, making the estimates less reliable.
%However, differences are present. 
These differences may have clinical or SE implications.
Thus, we do not claim that our synthetic data are clinically valid. 
Rather, our analyses illustrate that a generator can appear statistically plausible while still requiring domain-specific validation before use.
%Also, the other analysis provided us with valuable insights into the clinical validity of our dataset.
Our preliminary experimentation confirms that engineering these systems is complex: It is not a mere data-analytic problem; it is a SE problem that involves multiple dimensions and stakeholders, where the properties P to be considered for the validation of the dataset assume a primary value.

\section{Toward Property-Driven Synthetic Data Engineering}
\label{sec:challenges}

%In the following, we detail the \emph{lessons} we learned during our preliminary experimentation and the \emph{challenges} for the SE community we identified.
%Although our preliminary results were promising, we learned many lessons and identified several challenges for the software engineering community, which we share in this reflection paper. 

Our experience suggests that the central challenge is not choosing a better synthetic data generator, but engineering the validity conditions under which generated data can be trusted for a particular software task. 
We therefore frame SDG as a \emph{property-driven} engineering process. 
In this process, stakeholders define the properties that synthetic data must satisfy; developers operationalize those properties as checks; tools compare generators and configurations against those checks; and pipelines evolve as data, tasks, and regulations change.
This process provides an additional opportunity for automation for software engineers, i.e, to support developers in moving from informal stakeholder needs to executable validity checks.
We report a possible mapping between stakeholders, properties, and automation opportunities in \Cref{tab:property}.
\begin{table*}[]
    \centering
    \caption{Mapping of stakeholders to properties, checks and automation opportunities}
    \label{tab:property}
    \vspace{-.8em}
    \small
    \begin{tabular}{lp{.16\textwidth}p{.36\textwidth}p{.28\textwidth}}
    \toprule
         \textbf{Stakeholder} & \textbf{Desired property} & \textbf{Example property/check} & \textbf{Automation opportunity} \\ \midrule
         Clinician & Clinical plausibility & No impossible treatment/value combinations; survival curves clinically plausible & Constraint mining; rule checking; clinician-in-the-loop validation\\
         Data scientist & Statistical fidelity & Marginal distributions, correlations, task utility & Metric selection and model comparison\\
         Privacy officer & Disclosure protection & Low membership-inference risk; no near-duplicate patients & Privacy risk analysis; synthetic-data release gates\\
         Software Engineer & Test adequacy/Pipeline reliability & Synthetic data exercises target scenarios and edge cases & Property-based test generation; regression testing of data pipelines\\
         Maintainer & Evolution robustness & New data or hospital source does not invalidate properties & Monitoring; drift detection; impact analysis\\\bottomrule
    \end{tabular}
    \vspace{-.5em}
\end{table*}

Thanks to our preliminary experimentation, we learned the following \emph{lessons} (L):

\smallskip

\textbf{L1. Data cleaning is engineering, not pre-processing}.
The removal of variables and rows changes what properties can be preserved and what conclusions can be drawn:
Data cleaning decisions become part of the system specification because they determine which properties can later be synthesized, validated, and trusted.
%Limited and sensitive data fundamentally affect validation. The software engineering (and AI) community typically assumes that wide amounts of data characterize current applications. However, many applications lack sufficient data, and the available data are often sensitive. Many other domains and applications shar the problem considered in this reflection paper. For example, another case from the medical domain comes from a recent study that considered a data set of 1052 patients diagnosed with malignant neoplasm of female breast~\cite{zazzetti2025longitudinal}. Furthermore, our experience highlighted a critical SE challenge: real-world data are often not readily usable to support SE activities such as modeling, analysis, or automated data generation. Significant manual effort was required to clean, validate, and standardize the dataset before it could be reliably used. Such activities are not merely pre-processing steps, but essential engineering tasks that directly affect the correctness, reliability, and usefulness of software artifacts.

\textbf{L2. Synthetic data validity is multi-property}.
Statistical similarity, clinical plausibility, privacy, and task utility may conflict.
%In many practical situations, data contains extremely sensitive information. Our data contains personal information. Accessing this data is complicated, as it is necessary to avoid improper use and disclosure. For our study subject, accessing this data required significant time. The ethics committee of our partner hospital must approve any use of the provided data, including the publication of related results. 

\textbf{L3. There is no universal best generator}.
The choice of generator depends on data characteristics and stakeholder properties.
%A key challenge is the definition of the properties $P$ that synthetic data should satisfy. Different stakeholders impose different requirements: clinicians focus on clinical plausibility, data scientists on statistical fidelity, and legal stakeholders on privacy preservation. These properties are often conflicting, and improving one (e.g., accuracy) may negatively impact another (e.g., privacy). A recent work proposed the Synthetic Validation Framework~(SAFE)~\cite{10.1182/blood-2025-396} powered by Train~\cite{trainai_company} to evaluate the fidelity, utility, and privacy of synthetic data. Synthetic data were tested across three settings: (1)~integration with i2b2 for privacy-preserving data sets; (2)~multistate disease modeling to predict clinical outcomes; and (3)~generation of synthetic control groups for clinical trials. These studies show that generating high-quality synthetic data remains challenging. For example, a key challenge is balancing utility and privacy: Synthetic data should be realistic enough to be useful but different enough to avoid any link to the original data. Indeed, they should avoid exposing any sensitive information, as this would defeat its purpose.

\textbf{L4. Synthetic-data pipelines evolve}. 
As new patients, variables, hospitals, and tasks appear, %validity 
properties must be revisited.
%There are many implementations and alternatives for the synthetic data generation procedure ($f$). However, these alternatives are typically designed to target specific problems and do not distinguish between properties of interest for different stakeholders. For example, a recent work~\cite{zazzetti2025longitudinal} analyzed the effectiveness of  generative adversarial networks (GANs), variational autoencoders (VAEs), and language models (LMs) to synthesize patients diagnosed with malignant neoplasm of the female breast (ICD-9). While the authors showed the effectiveness of GANs within this context, our preliminary experiments revealed that GANs were not sufficiently effective in our context, and that TVAE provided much more reliable results.

%Considering all these lessons, we believe that there is a need for more effective software engineering solutions that can support software design within a context characterized by the absence of data and constrained by sensitive information.
%Specifically, we characterized the software engineering challenges by considering eight software engineering areas from the ASE 2026 call for papers~\cite{ASE2026}.
%https://conf.researchr.org/track/ase-2026/ase-2026-research-track

\smallskip

Based on these lessons, we identify three key SE \emph{challenges}.

\smallskip

%\claudio{Until Here.}

\emph{Requirements under uncertainty}. 
Precisely understanding and eliciting the requirements of the system is challenging.
Data may be used by different stakeholders (e.g., oncologists, other physicians, data managers, or legal offices), each of them with different expectations and requirements.
For example, oncologists are interested in the clinical validity of the generated data; data managers are concerned about statistical validity; legal offices need to ensure the confidentiality of personal information. 
Designing solutions that meet all of these requirements is complicated (e.g., we experienced different tradeoffs between clinical validity and confidentiality).
%We also noticed that solutions applicable in one domain may not apply to another.
%\andrea{Should we say that our formal framework allows for taking into account this difference in requirements thanks to the properties P?}
%parte di formal
A further challenge is the formalization of these requirements. Providing a precise mathematical definition of the properties $P$ is difficult, and small variations in their definition may significantly affect the evaluation of different synthesis solutions.
Future tools could mine candidate properties from real data, encode clinical and privacy constraints, detect property conflicts, recommend appropriate synthetic data generators, and continuously monitor whether a synthetic-data pipeline remains valid as the real dataset evolves.

\emph{Validation without ground truth}.
To be valuable for the SE community, synthetic data must preserve specific characteristics that enable obtaining valuable information from their analysis.
For example, in the absence of sufficient real data, it is difficult to define ground truth for validation.
Privacy constraints prevent cross-institution validation, forcing synthetic data to act as a surrogate for ground truth, which fundamentally alters validation pipelines.
We evaluated the generated data by comparing distributions and correlations to ensure that key statistical properties were preserved and performed clinical validation using survival curves~\cite{Bland1998}. 
% These statistical measures do not necessarily capture clinical validity.
%parte di sw analytics
From the software analytics perspective, defining appropriate metrics for assessing the quality of the generated data is a challenge. A single standard metric does not exist; Instead, multiple measures are required, including both statistical evaluation and clinical validation. However, these measures may not agree, as statistical similarity does not necessarily imply clinical validity.
%parte di AI & SE
Moreover, the characteristics of the data (e.g., heterogeneity, missing values, and imbalance) further complicate the evaluation, as different generation models may behave inconsistently across different types of variables.

\emph{Evolution of synthetic pipelines}. 
Data evolves as new patients are analyzed. 
(Synthesized) Data from other hospitals may need to be integrated.
Also, the pathology and the population characteristics change over time (e.g., average age).
Thus, new properties may be of interest for stakeholders.
Synthetic data pipelines must adapt to these changes.
In our context, some variables were removed due to missing values, but they may become relevant again as new data becomes available. This requires continuous adaptation of the synthesis process and validation criteria.

\definecolor{mediumelectricblue}{rgb}{0.01, 0.31, 0.59}
\definecolor{mayablue}{rgb}{0.45, 0.76, 0.98}

%\begin{figure}[t]
%\centering
%\resizebox{\columnwidth}{!}{
%\begin{tikzpicture}
%  \path[mindmap,concept color=mediumelectricblue,text=white]
%    node[concept] {Synthetic Data Generation}
%    [clockwise from=0]
%    child[concept color=mayablue!80!black] {
%      node[concept] {Data}
%      [clockwise from=90]
%      child[concept color=mayablue!90!black] { node[concept] {Small Datasets/Limited Data} }
%      child[concept color=mayablue!90!black] { node[concept]{Missing Values} }
%      child[concept color=mayablue!90!black] { node[concept] {Heterogeneity} }
%      child[concept color=mayablue!90!black] { node[concept] {Class Imbalance} }
%    }  
%    child[concept color=mayablue!80!black] {
%      node[concept] {Privacy}
%      [clockwise from=-30]
%       child[concept color=mayablue!90!black] { node[concept] {Re-identification Risk} }
%       child[concept color=mayablue!90!black] { node[concept] {Data Sharing Constraints} }
%    }
%    child[concept color=mayablue!80!black] {
%      node[concept] {Quality}
%      child[concept color=mayablue!90!black] { node[concept] {Statistical Fidelity} }
%      child[concept color=mayablue!90!black] { node[concept] {Different Applications} }
%    }
%    child[concept color=mayablue!80!black,grow=200] {
%      node[concept] {Validation}
%      [clockwise from=210]
%      child[concept color=mayablue!90!black] { node[concept] {Clinical Realism} }
%      child[concept color=mayablue!90!black] { node[concept] %{Evaluation Metrics} }
  %  };
%\end{tikzpicture}
%}
%\caption{Software Engineering Challenges.}
%\label{fig:mindmapFig}
%\end{figure}

\section{Related Work}
\label{sec:related}

Recent work has explored SDG techniques for tabular and medical datasets, focusing on improving data availability while preserving statistical and clinical properties. Zazzetti et al.~\cite{zazzetti2025longitudinal} propose a framework for generating longitudinal synthetic breast cancer data using models such as GANs, VAEs, and language models, and introduce a comprehensive validation pipeline (SAFE) to assess fidelity, privacy, and clinical utility. They demonstrate how synthetic data can support clinical research, including predictive modeling and the creation of synthetic control groups.
%In this work, we address a similar problem, focusing on the generation of synthetic tabular data in a setting characterized by limited data availability and strong privacy constraints.
%We provide a structured analysis of the main challenges that arise in this context, highlighting key issues related to data characteristics, model selection, and evaluation.
More broadly, several studies analyze generative approaches for tabular data, including GAN-based and VAE-based methods~\cite{10101315,shi2025synthetic}. 
These techniques aim to reproduce statistical patterns in the original data and are often evaluated through metrics such as distribution similarity, correlation preservation, and downstream task performance. 
Other work highlights the role of synthetic data in addressing data scarcity, imbalance, and privacy constraints in machine learning applications~\cite{Alzubaidi2023A,Goyal2024A}. 
While these approaches provide effective generation techniques and validation pipelines, they typically rely on predefined metrics or domain-specific evaluation procedures.
%
%
%Alzubaidi et al.~\cite{Alzubaidi2023A} studied how to train deep learning models when datasets are small, imbalanced, or not fully representative. In real-world applications, the quality of the data strongly affects model performance. One common solution is to use synthetic data generation to augment and re-balance datasets by creating additional high-quality samples. This approach helps reduce the problem of minority classes and improve the overall reliability of the models.
%\cite{Goyal2024A} analyzes synthetic data generation techniques based on generative AI, including large language models (LLMs), GANs, and VAEs. The study presents synthetic data as a solution to common challenges in machine learning, such as data scarcity, privacy concerns, and algorithmic bias.
%This work highlights several practical limitations, including high computational costs, training instability, and insufficient privacy guarantees, which limit their adoption in real-world scenarios.
%This work is closely related to our approach, as it addresses the same goal of using AI-based techniques for synthetic data generation. In particular, it evaluates models such as GANs and TVAE, which are also explored in our study. Moreover, it discusses challenges that are central to our work, including privacy preservation and the difficulty of training models on incomplete or limited datasets.

Synthetic data has also been widely used in SE to support testing~\cite{10.1145/3691620.3695067}, benchmarking, and empirical studies. Prior work uses synthetic datasets to evaluate security software, perform statistical testing~\cite{8115698}, and support experimental reproducibility when real-world data are scarce or proprietary~\cite{Yan2022}. In these contexts, synthetic data is treated as a practical substitute for real data~\cite{10.1145/3691620.3695058}, enabling controlled experimentation and scalability of evaluations.

%Unlike existing work, our paper outlines software engineering challenges. 
%It advocates the need for software engineering solutions that support the development of software systems in the absence of sufficient, shareable, representative, or task-specific data.
%It argues for and supports this need by reporting on our experience with a case study from the medical domain. 
In contrast, our focus is not on proposing a new generator or validation metric. We argue that SDG introduces an SE %lifecycle 
problem: stakeholders must define validity properties, developers must operationalize them into checks, tools must reason about trade-offs and conflicts, and pipelines must evolve as data and use cases change.
%Overall, these works show that data scarcity in safety-critical medical systems is still an important and not fully solved problem. While synthetic data generation has made progress, it is still difficult to train reliable models when data is limited and sensitive. Many existing approaches assume large and well-prepared datasets, which are often not available in real medical scenarios.

%\andrea{To expand}
\section{Conclusion and Future Work}
\label{sec:conclusions}
This paper reflects on our experience engineering software systems in a data-scarce and privacy-constrained medical domain for supporting oncologists of a major hospital. 
Our results suggest that data scarcity does not simply limit existing approaches but fundamentally challenges core assumptions of SE. 
In such settings, developers cannot rely on large datasets for validation or model construction. 
Instead, they must reason in terms of properties, constraints, and stakeholder-specific requirements.
Our preliminary evaluation shows that SDG is not merely a technical solution to data unavailability, but a complex engineering problem 
%The key challenge is no longer generating data, but determining which data generation technique to chose, which properties should be preserved, and how they can be validated in the absence of ground truth.
which introduces the necessity of considering statistical fidelity, clinical plausibility, and privacy, requiring explicit trade-offs that current methodologies do not adequately support or address.

Our observations highlight a broader shift for the SE community working in data-scarce and data-sensitive domains: from data-driven to property-driven approaches. 
Validation, requirements engineering, and system evolution must be rethought to explicitly account for limited, sensitive, and evolving data. 
We therefore argue that future research should focus on methods for defining, balancing, and validating properties in synthetic data. %, as well as on engineering practices for maintaining adaptive and trustworthy data generation pipelines. 
More broadly, SE must move beyond the assumption of abundant data and develop foundations for building and validating systems in its absence.

\section*{Data Availability Statement}
Due to the sensitivity of the dataset, neither the original nor the synthesized data can be shared at this time, as disclosure risks cannot yet be fully excluded. %However, the paper reports aggregated results sufficient to support our findings.

\bibliographystyle{ACM-Reference-Format}
%\balance
\bibliography{bibliography}

\end{document}